\documentclass[journal,comsoc]{IEEEtran}
\IEEEoverridecommandlockouts

\usepackage{amssymb}
\usepackage{lettrine}
\newtheorem{lemma}{Lemma}
\usepackage{graphicx}
\usepackage{epstopdf}

\usepackage[normalem]{ulem}
\usepackage{flushend}
\abovecaptionskip \belowcaptionskip

\usepackage{cite}

\ifCLASSINFOpdf

\else

\fi

\usepackage{amsmath}

\usepackage{algorithm}
\usepackage{algorithmicx}
\usepackage{algpseudocode}

\usepackage{array}

\usepackage{flushend}%\setlength{\abovedisplayskip}{4.8pt}

\usepackage{float}
\usepackage{color}
\usepackage{makecell}
\usepackage{multirow}
\usepackage{graphicx}
\usepackage{subfigure}
\usepackage{times}
\usepackage{array}

\begin{document}
\pagestyle{empty}
%
% paper title
% Titles are generally capitalized except for words such as a, an, and, as,
% at, but, by, for, in, nor, of, on, or, the, to and up, which are usually
% not capitalized unless they are the first or last word of the title.
% Linebreaks \\ can be used within to get better formatting as desired.
% Do not put math or special symbols in the title.
\title{Artificial-Noise-Aided Secure Channel with a Full-duplex Source}
%
%
% author names and IEEE memberships
% note positions of commas and nonbreaking spaces ( ~ ) LaTeX will not break
% a structure at a ~ so this keeps an author's name from being broken across
% two lines.
% use \thanks{} to gain access to the first footnote area
% a separate \thanks must be used for each paragraph as LaTeX2e's \thanks
% was not built to handle multiple paragraphs
%

\author{Xinyue Hu, Caihong Kai, Shengli Zhang, Zhongyi Guo and Jun Gao}
\author{Xinyue Hu $^{\dagger}$ Caihong Kai $^{\dagger}$ Shengli Zhang $^{\ast}$ Zhongyi Guo $^{\dagger}$ and Jun Gao $^{\dagger}$\\
$^{\dagger}$School of Computer Science and Information Engineering, Hefei University of Technology, Hefei, China\\
$^{\ast}$School of Information Engineering, Shenzhen University, Shenzhen, China\\
Email: huxinyue@mail.hfut.edu.cn, chkai@hfut.edu.cn, zsl@szu.edu.cn, {guozhongyi, gaojun}@hfut.edu.cn
}
\maketitle
\thispagestyle{empty}
% As a general rule, do not put math, special symbols or citations
% in the abstract or keywords.
\begin{abstract}
This paper consider a new secure communication scene where a full-duplex transmitter (Alan) need to transmit confidential information to a half-duplex receiver (Bob), with a silent eavesdropper (Eve) that tries to eavesdrop the confidential information. For realizing secure communication between Alan and Bob, a novel two phases communication scheme is proposed: in Phase 1, Alan and Bob send artificial noises (AN) simultaneously, while in Phase 2, Alan superimposes the AN received in Phase 1 with its confidential signal and sends the mixed signal to Bob. Since the mixed AN could degrade the SINR (Signal to Interference and Noise Ratio) of Eve, but does not affect the SINR of Bob, a secrecy capacity can be achieved. We also derive the conditions that the secrecy capacity of the proposed scheme exists, and analyze the secrecy outage probability under Rayleigh fading channel. Numerical results show that the secrecy capacity is about two times higher than without AN, even though in the proposed scheme half of the time is used to transmit ANs, and the outage probability is about five times lower than that without AN.
\end{abstract}

% Note that keywords are not normally used for peerreview papers.

% For peer review papers, you can put extra information on the cover
% page as needed:
% \ifCLASSOPTIONpeerreview
% \begin{center} \bfseries EDICS Category: 3-BBND \end{center}
% \fi
%
% For peerreview papers, this IEEEtran command inserts a page break and
% creates the second title. It will be ignored for other modes.
\IEEEpeerreviewmaketitle

\section{Introduction}
Due to the broadcast nature of wireless channels, wireless security is an important concern that is attracting increasing interests from the community. Traditionally, the security problem in wireless networks was mainly studied at higher layers using key-based cryptographic methods. However, as the computational capability of wireless devices grows rapidly, perfect security can be hardly guaranteed with the key-based solutions. Physical-layer communication security that was first introduced by Shannon \cite{1} emerges as an effective solution for secure communications by exploiting the physical characteristics of wireless channels. In \cite{2}, Wyner showed that a transmitter can communicate to its receiver with perfect secrecy from an information theoretic perspective, when the eavesdropper channel (between the transmitter and the eavesdropper) is degraded with respect to the main channel (between the transmitter and the receiver). Based on this result, various methods have been proposed to improve secrecy capacity by means of degrading the eavesdropper channel, and one important direction is to inject artificial noises (AN) to degrade the eavesdropper channel, such as \cite{3,4,5,6,7,8,9,10}. Specifically, the confidential information is superimposed with some specially-designed artificial noises, which can be canceled by the receiver while remains an interference to the eavesdropper.

More specifically, in \cite{3,4,5,6}, the AN is a special signal designed in the null space of the main channel, and is emitted to interfere with the eavesdropper. That is, in \cite{3,4,5,6}, the MIMO (Multiple-Input Multiple-Output) equipment and the beamforming technology are needed, further, for playing beamforming, CSI (Channel State Information must be known by the node who playing beamforming. In \cite{7,8}, the AN generated by transmitter of friendly jammer is assumed pre-known by the receiver but unknown by the eavesdropper. Its means in \cite{7,8} the risk that the selected AN may be eavesdropped by the eavesdropper is existing. Recently, full-duplex technology has been incorporated to realize physical-layer secrecy communication \cite{9,10}, in which the authors considered a scenario in which either the relay node or the receiver has full-duplex communication ability. Thus, the relay or receiver can receive the information signal and transmit AN simultaneously to puzzle the eavesdropper, thus the AN-leakage problem in \cite{9,10} can be avoided.

Different from \cite{9,10}, this paper consider a new secure communication scene where a full-duplex transmitter (Alan) need to transmit confidential information to a half-duplex receiver (Bob), with a silent eavesdropper (Eve) that tries to eavesdrop the confidential information, and to the best of our knowledge, there are no works have solved the problem that how to realize the secrecy transmission from a full-duplex transmitter to a half-duplex receiver without the help of relays. It is worth noting that the above problem is based on practical scene, for example, a full-duplex base station (BS) wants to transmit confidential information to a user while prevents other users from eavesdropping. To counter the above problem, a novel two-phase communication scheme that utilizes the advantages of full-duplex technology and AN is proposed: In phase 1, the transmitter and receiver send artificial noise ${n_A}$ and ${n_B}$ simultaneously (${n_A}$ and ${n_B}$ are only knows by Alan and Bob respectively), while in phase 2, Alan mixes the received signal with the confidential signal and sends it to Bob, thanks for full-duplex, the received signal in phase 1 at Alan is mainly constituted by ${n_B}$. After phase 2, because ${n_B}$ is known by Bob, so Bob could cancel the AN from the received signal, but the eavesdropper (Eve) only knows ${n_B}$ polluted by ${n_A}$ from phase 1, so Eve could not cancel the AN from received signal, then Eve must suffer more interference than Bob. We note that throughout the whole transmission process, both CSI and the artificial noise sent by Bob in phase 1 are not required by Alan, Alan also does not need the assistance of helpers or relays and does not need to perform beamforming, what Alan need to do, is just only mix and forward. Although this scheme only uses half times to transmission, but it still can prominent improve the secrecy capacity and reduce the secrecy outage probability than that without AN and full-duplex.

The contributions of this paper are summarized as follows:
\begin{itemize}
\item We propose a new transmission scheme to realize secure communication from full-duplex transmitter to half-duplex receiver. This scheme is simple to implement.
\item We drive the conditions that positive secrecy capacity of the proposed scheme exists, analyze the secrecy outage probability and the ergodic secrecy capacity under Rayleigh fading channel.
\end{itemize}

The remainder of the letter is organized as follows: Section \uppercase\expandafter{\romannumeral2} describes the considered system model and introduces the proposed scheme. Section \uppercase\expandafter{\romannumeral3} analyzes the secrecy performance of the proposed scheme and gives the secrecy capacity and outage probability of the proposed scheme. Section \uppercase\expandafter{\romannumeral4} presents the numerical results and finally Section \uppercase\expandafter{\romannumeral5} concludes the paper.

\section{System Model and the Proposed Scheme}
This section presents our system model and the proposed secure communication scheme.

\subsection{System Model}
As shown in Fig. 1, we consider a wireless communication system with three communication nodes: a full-duplex transmitter (Alan), an intended receiver (Bob) with single-antenna and a silent eavesdropper (Eve) with single-antenna. Alan need to send confidential information to Bob, and Eve tries to eavesdrop the information.
\begin{figure}[H]
  \centering
  % Requires \usepackage{graphicx}
  \includegraphics[width=2.3in]{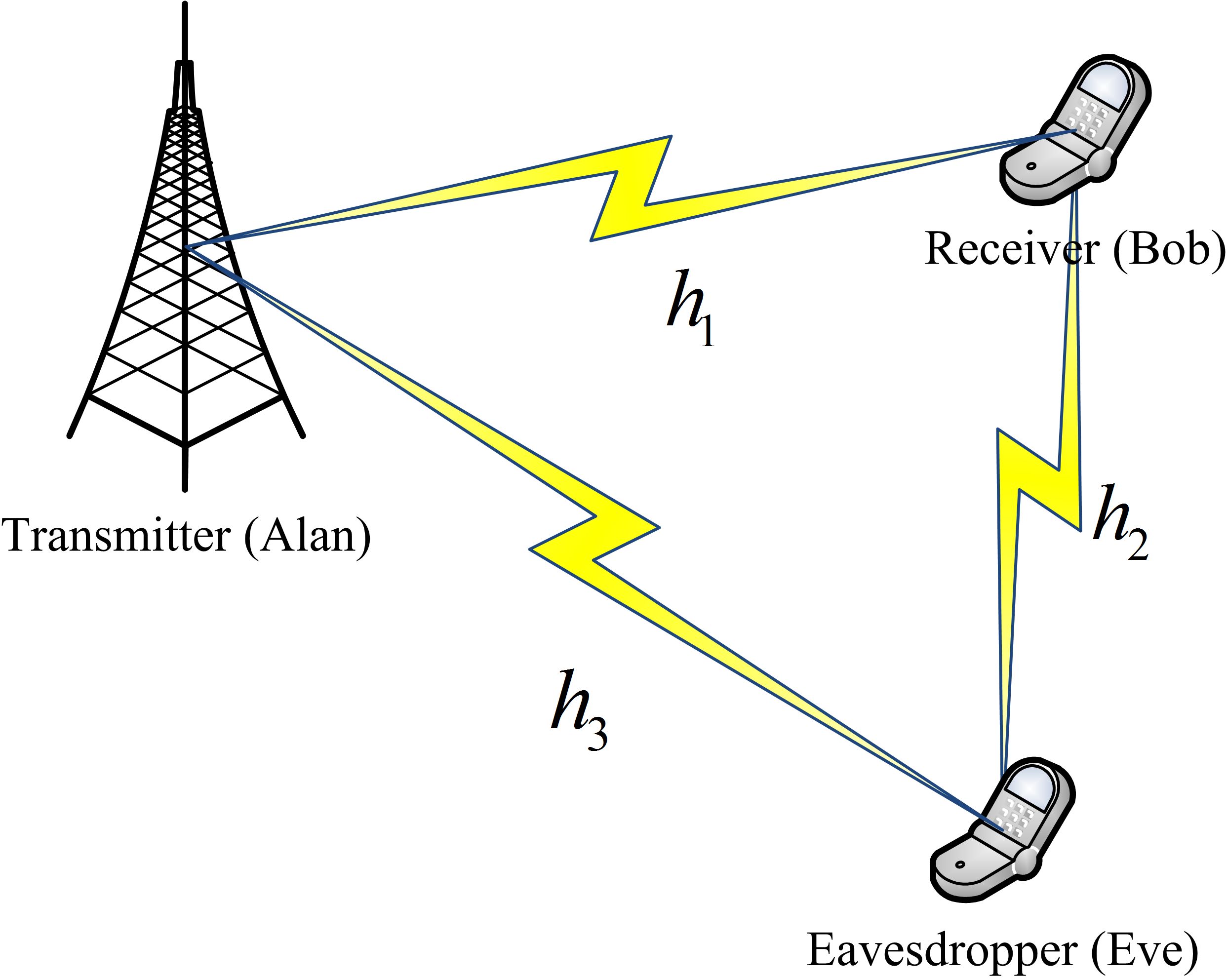}\\
  \caption{The Considered Communication System.}\label{1}
\end{figure}

Let ${h_1}$ and ${h_3}$ be the complex channel fading coefficient of the main channel and the eavesdropper channel, respectively, and ${h_2}$ be the complex channel fading coefficient between Bob and Eve. We assume that at the initial phase of secrecy communication, all the channel coefficients are unknown to Alan, Bob and Eve. Moreover, we assume the reciprocity of the forward and backward channels and a block-fading channel model where each channel coefficient remains unchanged for the time slot duration of $T$ seconds is assumed in this paper\footnote{Here we make this assumption for the purpose of easy analysis. In fact, the assumption of the reciprocity channel is not necessary. Bob could use the proposed method same as Eve (as in Lemma 1 ) to cancel the interference caused by the artificial generated by itself.}. Let the AWGN (Additive Gaussian White Noise) at each node has the same power ${N_0}$.

\subsection{The Proposed Scheme}

For guaranteeing the confidential information transmitted from Alan to Bob, a two-phase communication scheme is proposed.

\textbf{Phase 1:} In the first phase, which lasts for $T/2$ seconds, Bob sends an artificial Gaussian noise $\sqrt {{P_B}} {n_B}$ (note that ${n_B}$ is only known by Bob) in which ${n_B}{\rm{\sim}}{{\cal C}{\cal N}}(0,1)$ and ${P_B}$ is the transmit power of Bob. When Bob sends $\sqrt {{P_B}} {n_B}$, Alan synchronously sends another artificial Gaussian noise $\sqrt {{P_A}} {n_A}$ (note that ${n_A}$ is only known by Alan) , in which ${n_A}{\rm{\sim}}{{\cal C}{\cal N}}(0,1)$ and ${P_A}$ is the transmit power of Alan.

After Phase 1\footnote{For synchronizing, in phase 1, a pilot signal is added ahead $\sqrt {{P_B}} {n_B}$, so after phase 1, Alan knows ${h_1}$ and Eve knows ${h_2}$.}, denote the signal received by Alan (Thanks to the full-duplex ability, Alan can transmit and receive simultaneously.) and Eve by ${y_A}$ and ${y_{E1}}$, respectively. We have:
\begin{equation}
{y_A} = {h_1}\sqrt {{P_B}} {n_B} + \sqrt {\lambda {P_A}} {n_0} + {n_{A1}},
\end{equation}
\begin{equation}
{y_{E1}} = {h_2}\sqrt {{P_B}} {n_B} + {h_3}\sqrt {{P_A}} {n_A} + {n_{E1}},
\end{equation}
 where $\sqrt {\lambda {P_A}} {n_0}$ is caused by the residual self-interference (RSI) in which ${n_0}{\rm{\sim}}{{\cal C}{\cal N}}(0,1)$ and $\lambda  = {\left| {{h_{SI}}} \right|^2}$ is the RSI channel gain of Alan, and indicates the self-interference (SI) cancellation capability of Alan, where ${h_{SI}}$ is the RSI channel of Alan. Note that RSI=0 denotes perfect cancellation capability. ${n_{A1}}$ and ${n_{E1}}$ are AWGN satisfying ${{\cal C}{\cal N}}(0,{N_0})$, respectively.

Note that, the residual self-interfering channel gain $\lambda$ is determined by the applied SI cancellation algorithm. Here, we consider the digital-domain cancellation, where ${h_{SI}}$ can be presented as ${h_{SI}} = {h_{S{I_c}}} - {\hat h_{S{I_c}}}$ where ${h_{S{I_c}}}$ and ${\hat h_{S{I_c}}}$ are the self-interfering channel and its estimate value, this allows $\lambda$ to be modeled as a constant value \cite{11}.

\textbf{Phase 2:} In the second phase, which lasts for $T/2$ seconds,  Alan superimposes the received signal ${y_A}$ with the information signal $\sqrt {{P_s}} {s_A}$, where ${s_A}{\rm{\sim}}{{\cal C}{\cal N}}(0,1)$ is the confidential information signal and ${P_s}$ is the transmit power of ${s_A}$. The mixed signal is ${x_A} = \sqrt {{P_s}} {s_A} + {y_A}$. Alan then add a pilot signal ahead of ${x_A}$ and sent it to Bob(i.e., after Phase 2, $h_1$ and $h_3$ are known by Bob and Eve, respectively). Denote the received signal at Bob and Eve in Phase 2 by ${y_B}$ and ${y_{E2}}$, respectively. We have:
\begin{equation}
{y_B} = {h_1}\sqrt {{P_s}} {s_A} + {h_1}{h_1}\sqrt {{P_B}} {n_B} + {h_1}\sqrt {\lambda {P_A}} {n_0} + {h_1}{n_{A1}} + {n_{B2}},
\end{equation}
\begin{equation}
{y_{E2}} = {h_3}\sqrt {{P_s}} {s_A} + {h_3}{h_1}\sqrt {{P_B}} {n_B} + {h_3}\sqrt {\lambda {P_A}} {n_0} + {h_3}{n_{A1}} + {n_{E2}},
\end{equation}
where ${n_{B2}}$ and ${n_{E2}}$ are AWGN with variance $N_0$.

Since $\sqrt {{P_B}}{n_B}$  and $h_1$ are known by Bob, Bob could cancel the ${h_1}{h_1}\sqrt {{P_B}} {n_B}$ term from ${y_B}$ easily. By contrast, it is difficult for Eve to detect ${n_B}$ from $y_{E1}$. Furthermore, $h_1$ is unknown to Eve.Thus, Eve could not cancel the ${h_1}{h_3}\sqrt {{P_B}} {n_B}$ term in $y_{E2}$. Thus, Eve suffers more interference from ANs than Bob, and the proposed two-phase transmission scheme achieves higher secrecy capacity.

\section{Analysis on Secrecy Capacity and Outage Probability}

We next analyze the secrecy capacity and outage probability of the assumed Rayleigh block fading channel.

\subsection{Capacity of the Main Channel}

Recall that a pilot signal is transmitted in Phase 2, Bob and Eve can estimate ${h_1}$ and ${h_3}$, respectively. For Bob, it knows $\sqrt {{P_B}} {n_B}$ and ${h_1}$. From $y_B$ in (3), it can cancel the interference caused by ${h_1}{h_1}\sqrt {{P_B}} {n_B}$ and obtain ${y_{BS}}$ :
\begin{equation}
{y_{BS}} = {h_1}\sqrt {{P_s}} {s_A} + {h_1}\sqrt {\lambda {P_A}} {n_0} + {h_1}{n_{A1}} + {n_{B2}}.
\end{equation}

Define ${g_1}{\rm{ = }}{\left| {{h_1}} \right|^2},{g_2}{\rm{ = }}{\left| {{h_2}} \right|^2}$ and ${g_3}{\rm{ = }}{\left| {{h_3}} \right|^2}$ as the  gain of the three channels, respectively. The SINR of the main channel is:
\begin{equation}
SIN{R_B} = \frac{{{g_1}{P_s}}}{{{g_1}\lambda {P_A} + {g_1}{N_0} + {N_0}}},
\end{equation}
and the channel capacity of Bob, $C_B$, is\footnote{In this paper, the bandwidth term in Shannon formula is ignored. That is, the unit of capacity calculated in this paper is ¡°bit/Hz/second¡± or ¡°bit/symbol¡±.}:
\begin{equation}
{C_B} = {\log _2}(1 + SIN{R_B}).
\end{equation}

\subsection{Capacity of the Eavesdropper Channel}
The signals received by Eve in two phases are (2) and (4), respectively. We assume Eve is smart enough and could use $y_{E1}$ in (2) to reduce the interference in $y_{E2}$ as much as possible.

For cancelling the interference, Eve multiplies $y_{E1}$ with a complex coefficient $h_x^{}$, then obtains $h_x^{}{y_{E1}}$, and  minus it from $y_{E2}$. After cancel the interference, the signal obtained by Eve is :
\begin{equation}
\begin{split}
{y_{ES}} &= {h_3}\sqrt {{P_s}} {s_A}\\
 &+ \underbrace {{h_3}{h_1}\sqrt {{P_B}} {n_B} + {h_3}\sqrt {\lambda {P_A}} {n_0} + {h_3}{n_{A1}} + {n_{E2}}}_N\\
 &- \underbrace {h_x^{}\left( {{h_2}\sqrt {{P_B}} {n_B} +{h_3}\sqrt {{P_A}} {n_A} + {n_{E1}}} \right)}_{h_x^{}{y_{E1}}}.
\end{split}
\end{equation}

As regarding to how $h_x$ is determined to maximize the SINR of Eve, we have the following lemma:
\begin{lemma}
To achieve the highest SINR, Eve can select
\begin{equation}
h_x^ *  = \frac{{{P_B}|{h_1}{h_2}{h_3}|}}{{{g_2}{P_B} + {g_3}{P_A} + {N_0}}}{e^{({\theta _1} + {\theta _3} - {\theta _2})i}}
\end{equation}
in which ${\theta _1}$, ${\theta _2}$ and ${\theta _3}$ are the phases of ${h_1}$, ${h_2}$ and ${h_3}$, respectively.
\end{lemma}
\begin{IEEEproof}
For achieving the highest SINR, $h_x^ *$ must be:
\begin{equation}
h_x^ *  = \mathop {\arg \max }\limits_{{h_x}} \left( {SIN{R_{{y_{ES}}}}} \right),
\end{equation}
in which
\begin{equation}
SIN{R_{{y_{ES}}}} = \frac{{{g_3}{P_s}}}{{\underbrace {Var\left( {N - h_x^{}{y_{E1}}} \right)}_{EN}}},
\end{equation}
and we have
\begin{equation}
\begin{split}
EN &= {\left| {{h_3}{h_1} - {h_x}{h_2}} \right|^2}{P_B} + {\left| {{h_x}} \right|^2}{g_3}{P_A} + {\left| {{h_x}} \right|^2}{N_0}\\
 &+ {g_3}\lambda {P_A} + {g_3}{N_0} + {N_0}.
 \end{split}
 \end{equation}
From (11), we have  $h_x^ *  = \mathop {\arg \min }\limits_{{h_x}} \left( {EN} \right)$. From (12), it is easy to see that when the phase of ${h_3}{h_1}$ is equal to that of ${h_x}{h_2}$, $\left| {{h_3}{h_1} - {h_x}{h_2}} \right|$ is the smallest. That is, $\theta _x^* = {\theta _1} + {\theta _3} - {\theta _2}$.

After $\theta _x^*$ is determined, we can write $EN$ as:
\begin{equation}
\begin{split}
EN &= \left( {{g_2}{P_B} + {g_3}{P_A} + {N_0}} \right){\left| {{h_x}} \right|^2} - 2\left| {{h_1}} \right|\left| {{h_2}} \right|\left| {{h_3}} \right|{P_B}\left| {{h_x}} \right|\\
 &+ {g_1}{g_3}{P_B} + {g_3}\lambda {P_A} + {g_3}{N_0} + {N_0},
 \end{split}
\end{equation}
which is a convex function of $\left| {{h_x}} \right|$ and it is easy to verify that$\left| {h_x^*} \right| = \frac{{{P_B}|{h_1}{h_2}{h_3}|}}{{{g_2}{P_B} + {g_3}{P_A} + {N_0}}}$.
\end{IEEEproof}

Lemma 1 above gives out the value of $h_x^{}$ to make the SINR of Eve highest, but in our system, ${h_1}$ (correspondingly ${g_1}$) is unknown for Eve, so Eve could not compute $h_x^ *$ directly. Eve could approximate achieve $h_x^ *$ buy using exhaust method: firstly, Eve could set $\left| {h_x^{}} \right| = 1$ and exhaust ${\theta _x}$ to achieve ${\theta _x} = \theta _x^* = {\theta _1} + {\theta _3} - {\theta _2}$ makes $EN$ smallest, secondly, Eve fixes ${\theta _x} = \theta _x^* = {\theta _1} + {\theta _3} - {\theta _2}$ and exhausts $\left| {h_x^{}} \right|$ to achieve $\left| {h_x^*} \right| = \frac{{{P_B}|{h_1}{h_2}{h_3}|}}{{{g_2}{P_B} + {g_3}{P_A} + {N_0}}}$ makes $EN$ smallest. From the above exhaust method, we can see that, if $T$ is long enough and the exhaust step length is small enough, Eve could achieve exact $h_x^*$ , because if $T$ is long enough, the sample variance of the received signal at Eve will be almost equal to the real variance of the received signal, and if the exhaust step length is small enough, Eve could achieve exact $h_x^*$ obviously. In the remainder of this letter, we think the exact $h_x^*$ can be achieved by Eve.

To ease expression, we define $M$ as:
\begin{equation}
M = \frac{{P_B^2{g_2}{g_3}}}{{{g_2}{P_B} + {g_3}{P_A} + {N_0}}},
\end{equation}
By submitting $h_x^*$ into (8), we get can the SINR of Eve:
\begin{equation}
SIN{R_E} = \frac{{{g_3}{P_s}}}{{{g_3}\lambda {P_A} + {g_3}{N_0} + {N_0} + {g_1}{g_3}{P_B} - {g_1}M}},
\end{equation}
and the channel capacity is:
\begin{equation}
{C_E} = {\log _2}(1 + SIN{R_E}).
\end{equation}

\subsection{Secrecy Capacity and Outage Probability}
According to [2], the instantaneous secrecy capacity is:
\begin{equation}
\begin{split}
{C_S} &= {\left[ {{C_B} - {C_E}} \right]^ + }\\
& = {\left[ {{{\log }_2}\left( {1 + SIN{R_B}} \right) - {{\log }_2}\left( {1 + SIN{E_E}} \right)} \right]^ + }.
\end{split}
\end{equation}

For expressing clearly, we define $Y = {P_B} - M/{g_3}$ and $Z = \lambda {P_A} + {N_0}$. To achieve a expected secrecy rate ${R_s}$ (i.e., ${C_B} - {C_E} \ge {R_s}$), we require
\begin{equation}
\frac{{1 + \frac{{{g_1}{P_s}}}{{{g_1}\lambda {P_A} + {g_1}{N_0} + {N_0}}}}}{{1 + \frac{{{g_3}{P_s}}}{{{g_3}\lambda {P_A} + {g_3}{N_0} + {N_0} + {g_1}{g_3}{P_B} - {g_1}M}}}} \ge {2^{{R_s}}},
\end{equation}

After mathematical derivation, we can obtain:
\begin{equation}
{g_1} \ge {g_{1L}}\left( {{g_2},{g_3},R_s} \right) \buildrel \Delta \over = \frac{{ - B + \sqrt {{B^2} - 4AC} }}{{2A}},
\end{equation}
where $A = \left( {1 - {2^{{R_s}}}} \right)YZ + Y{P_s}$, ${\rm{B}} = \left( {1 - {2^{{R_s}}}} \right)\left( {{Z^2} + Z{N_0}/{g_3} + Y{N_0} + Z{P_s}} \right) + {P_s}{N_0}/{g_3}$ and $C = \left( {1 - {2^{{R_s}}}} \right)\left( {Z{N_0} + N_0^2/{g_3}} \right) - {2^{{R_s}}}{P_s}{N_0}$. The above inequation is the condition that the proposed scheme could achieve the target secrecy rate ${R_s}$.

We assume the channels are Rayleigh fading then calculate the secrecy outage probability and ergodic secrecy capacity. We write ${h_j}\sim{{\cal C}{\cal N}}(0,\sigma _j^2)$, $j \in \left\{ {1,2,3} \right\}$, the PDF(Probability Density Function)of ${g_j},j \in \left\{ {1,2,3} \right\}$ is
\begin{equation}
P({g_j}) = \frac{1}{{\sigma _j^2\Gamma (1)}}{e^{ - \frac{{{g_j}}}{{\sigma _j^2}}}},j \in \left\{ {1,2,3} \right\}.
\end{equation}

The secrecy outage probability of the proposed scheme, ${P_{out}(R_s)} = P\left( {{g_1} < {g_{1L}}\left( {{g_2},{g_3},R_s} \right)} \right)$, can be computed by
\begin{equation}
\begin{array}{l}
{P_{out}}({R_s}) = \\
\int_0^\infty  {\int_0^\infty  {\int_0^{{g_{1L}}\left( {{g_2},{g_3},{R_s}} \right)} {{e^{ - \frac{{{g_1}}}{{\sigma _1^2}}}}} } } {e^{ - \frac{{{g_2}}}{{\sigma _2^2}}}}{e^{ - \frac{{{g_3}}}{{\sigma _3^2}}}}\frac{1}{{\sigma _1^2\sigma _2^2\sigma _3^2}}d{g_1}d{g_2}d{g_3},
\end{array}
\end{equation}
and the ergodic secrecy capacity is
\begin{equation}
\begin{split}
E({C_S}) = &\int_0^\infty  {\int_0^\infty  {\int_{{g_{1L}}\left( {{g_2},{g_3},0} \right)}^\infty  {{e^{ - \frac{{{g_1}}}{{\sigma _1^2}}}}} } } {e^{ - \frac{{{g_2}}}{{\sigma _2^2}}}}{e^{ - \frac{{{g_3}}}{{\sigma _3^2}}}}\bullet\\
&\frac{1}{{\sigma _1^2\sigma _2^2\sigma _3^2}}{C_S}({g_1},{g_2},{g_3})d{g_1}d{g_2}d{g_3}.
 \end{split}
\end{equation}
where ${C_S}({g_1},{g_2},{g_3})$ is a function of ${C_S}$ about ${g_1}$, ${g_2}$, and ${g_3}$ given in (17).

\section{Numerical Results}
This section presents numerical results to evaluate the performance of the proposed scheme. Fig. 2 shows the numerical results of secrecy capacity with different ${P_s}$, where ${P_A} = {P_B} = 200$, ${g_1} = 0.4,{\rm{ }}{g_3} = 0.6$, ${N_0} = 1$ and $\lambda  = 0.0001$. Since the channel fading is given and ${g_1} < {g_3}$ , the secrecy capacity with no full-duplex and AN is always zero. By contrast, our proposed scheme can achieve positive secrecy capacity and the outage probability is zero. As shown in Fig. 2, the secrecy capacity of the proposed scheme increases as ${P_s}$ increases. As ${g_2}$ increases, the achieved secrecy capacity decreases. That is because as ${g_2}$ increases, Eve could obtain more accurate ${n_B}$ from (2), then could cancel the noise term ${h_3}{h_1}\sqrt {{P_B}} {n_B}$ in (4) more effectively.

\begin{figure}[t]
  \centering
  % Requires \usepackage{graphicx}
  \includegraphics[width=2.6in]{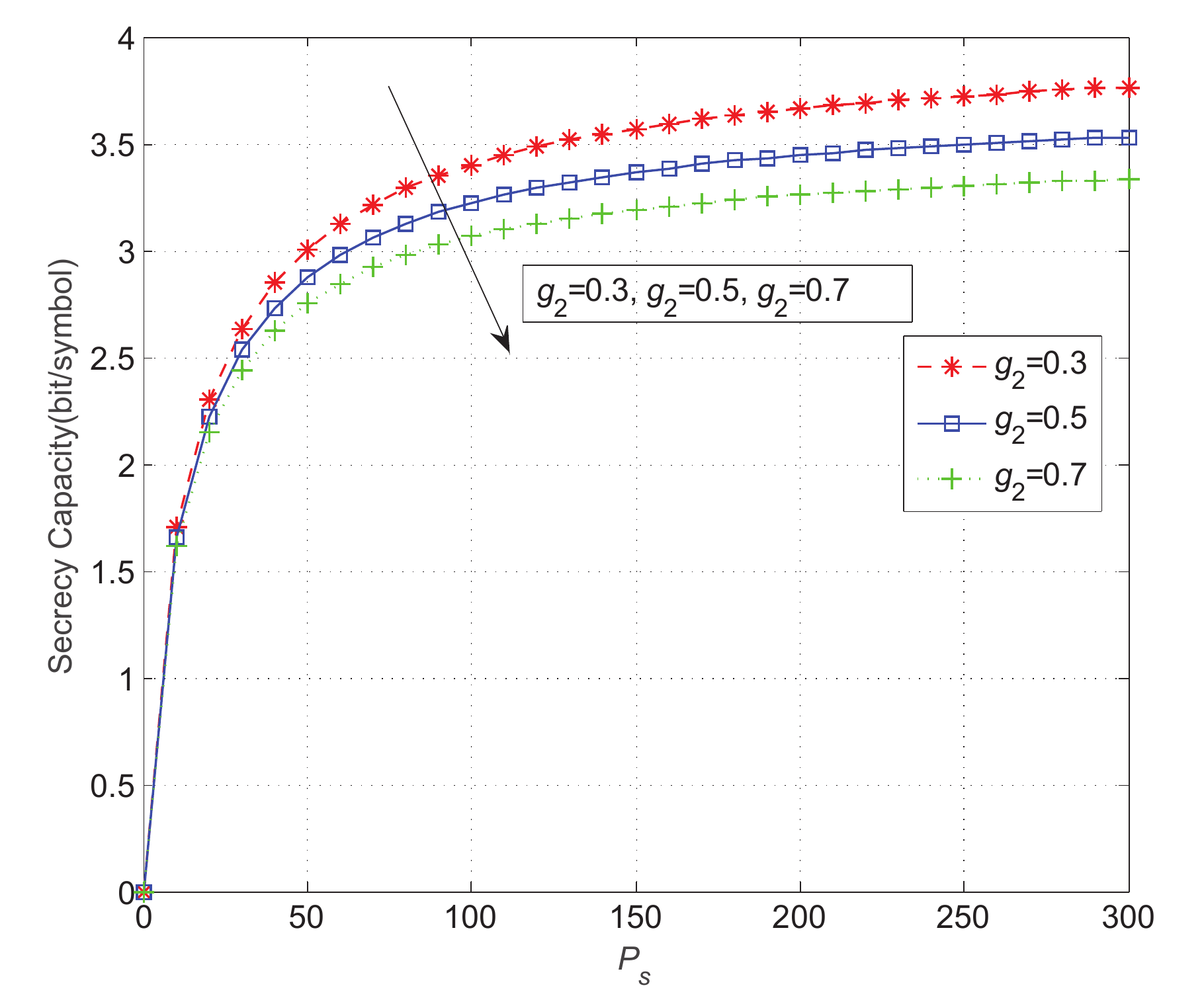}\\
  \caption{Secrecy capacity versus ${P_s}$,with different ${g_2}$, in which ${P_A}={P_B}=200$, ${g_1} = 0.4$ and ${g_3} = 0.6$.}\label{2}
\end{figure}

 Fig. 3 presents the numerical results of secrecy capacity with different ${P_{\rm{A}}}$ and ${P_{\rm{B}}}$, where ${P_s}=200$, ${g_1} = 0.4,{\rm{ }}{{\rm{g}}_2} = 0.7,{\rm{ }}{g_3} = 0.6$, ${N_0} = 1$ and $\lambda  = 0.0001$. As can be seen, as ${P_{\rm{A}}}$ and ${P_{\rm{B}}}$ increase, the secrecy capacity increases. That is because as the power of AN increases, Eve suffers from more interference.
 \begin{figure}[t]
  \centering
  % Requires \usepackage{graphicx}
  \includegraphics[width=2.6in]{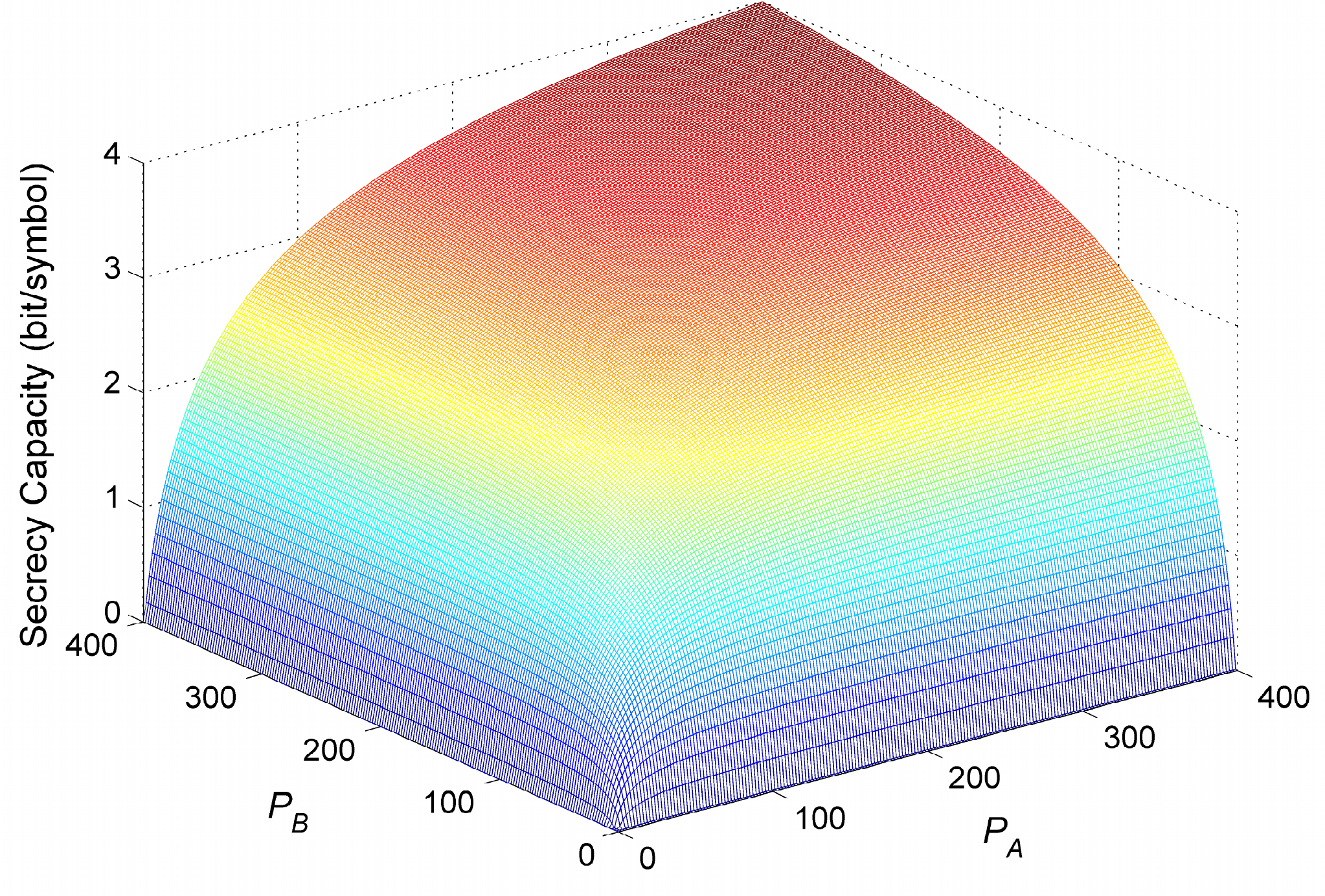}\\
  \caption{Secrecy capacity versus ${P_A}$ and ${P_B}$, in which ${g_1} = 0.4, {g_2} = 0.4,{\rm{ }}{g_3} = 0.6$, and ${P_s}=200$. }\label{3}
\end{figure}

For fading channels, we assume ${h_j}\sim{{\cal C}{\cal N}}(0,1),j \in \left\{ {1,2,3} \right\}$ and the other parameters are same as Fig. 2. Fig. 4 presents the secrecy capacities of the proposed scheme. Note that the ergodic secrecy capacity given by (22) is multiplied with 0.5 because half of the time is used to transmit ANs. As shown in Fig. 4, the proposed scheme achieves much higher secrecy capacity than the scheme without AN, even though in the proposed scheme half of the time is used to transmit random signals. On the other hand, as $\lambda$ increases, the ergodic secrecy capacity decreases, since a larger $\lambda$ corresponds to a severe residual noise caused by ${P_A}$ in Alan's full-duplex transmission. It is natural to see that more effective SI cancellation gives higher ergodic secrecy capacity. Also, in Fig. 4 if SI cancellation is bad, and the $P_s$ is small, the proposed scheme is worse than without AN, that is because the ratio of the noise caused by SI to $P_s$ is much bigger, but as long as the $P_s$ is not to small, the proposed scheme is still batter then without AN.
\begin{figure}[t]
  \centering
  % Requires \usepackage{graphicx}
  \includegraphics[width=2.6in]{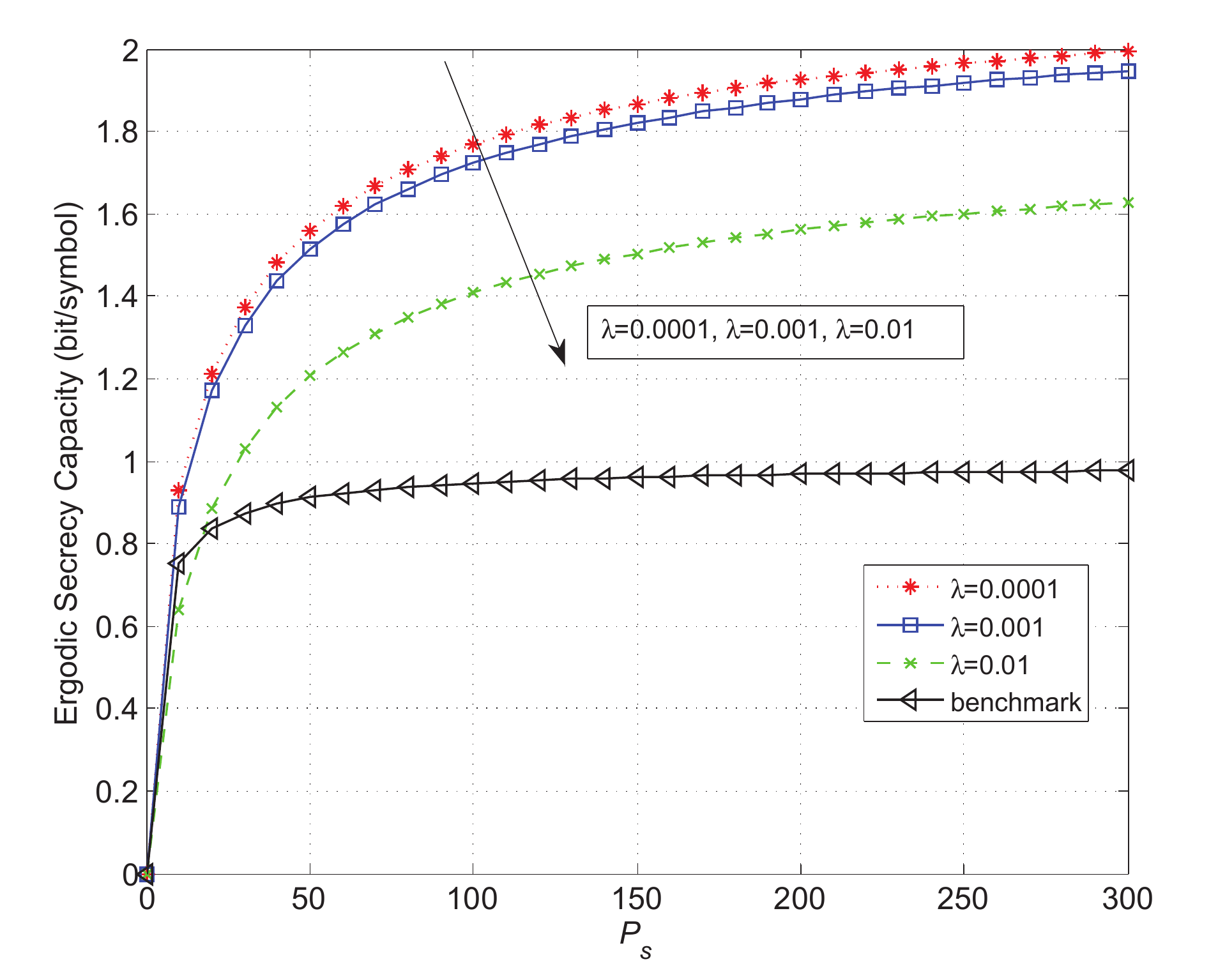}\\
  \caption{Ergodic secrecy capacity versus ${P_s}$, with different $\lambda $, in which ${P_A}={P_B}=200$. }\label{4}
\end{figure}

Fig. 5 shows the outage probability, where ${P_s}=400$, $P_A=P_B=P_{AN}$ and the other parameters are same as Fig. 3 used. As the target secrecy rate $R_s$ increases, the outage probability increases, but in our scheme, the outage probability is still lower than without AN. Also as the power of AN, $P_{AN}$, increases, the outage probability decreases, that is because the interference caused by AN at Eve become more effective.
\begin{figure}[t]
  \centering
  \includegraphics[width=3.2in]{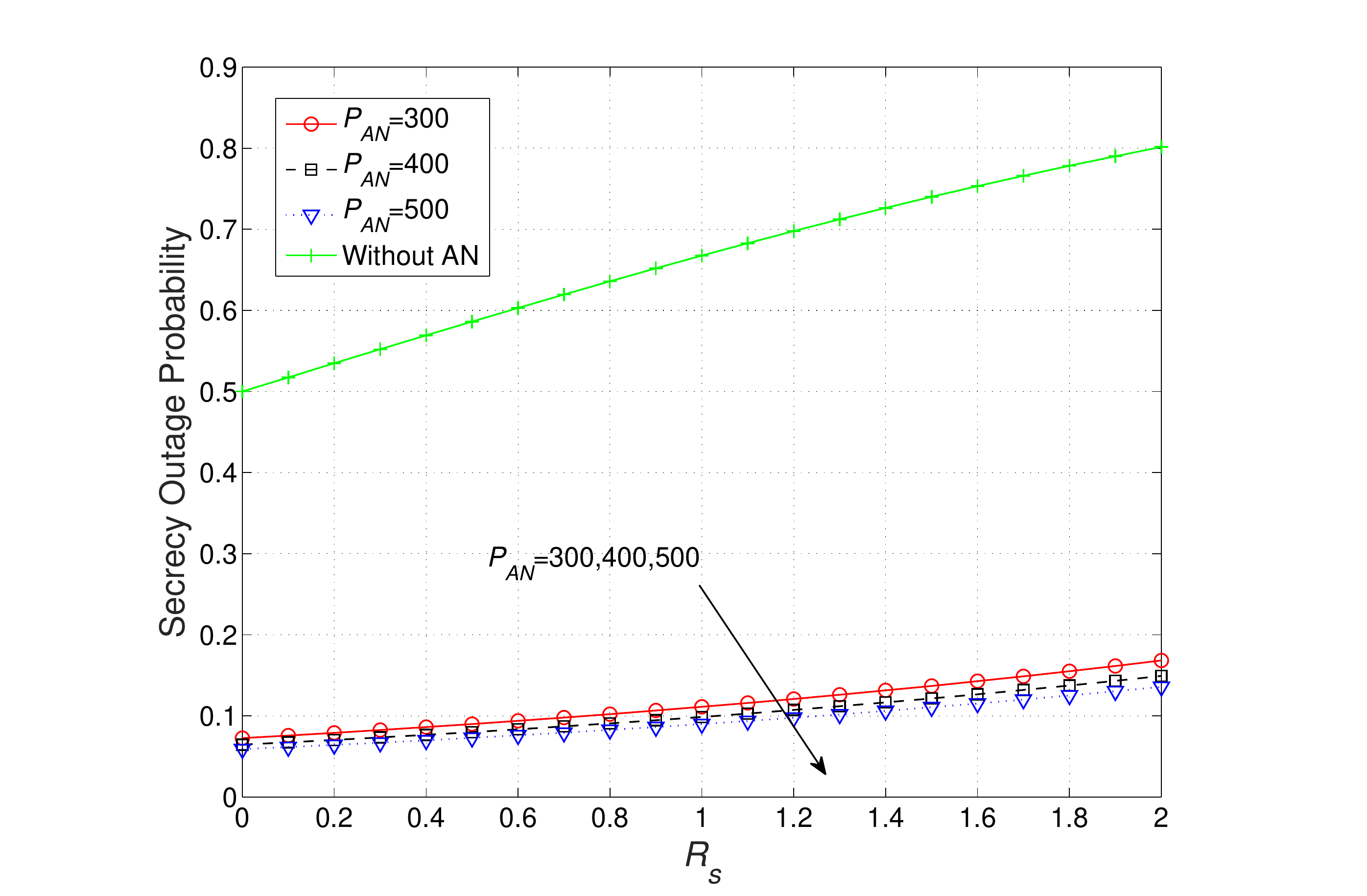}\\
  \caption{Outage Probability versus Transmit power of ANs.}\label{5}
\end{figure}
\section{Conclusions}
We proposed a two-phase transmission scheme to guarantee secrecy information transmission from a full-duplex source to a half-duplex destination. We analyzed the secrecy capacity of the proposed scheme, and presented the conditions for the proposed scheme to achieve positive secrecy capacity. Numerical results showed that the proposed scheme can obtain good performance in terms of both secrecy capacity and outage probability. In this paper, we assume all the ANs are transmitted with the same power. As a future work, we will investigate the power allocation algorithm of ANs and the secrecy performance of the proposed scheme could be further improved.
\bibliographystyle{IEEEtran}
\bibliography{refan}
% that's all folks
\end{document}